\documentclass[12pt]{article}
\usepackage{graphicx, epsfig}
\usepackage{amsmath}
\usepackage{latexsym}
\usepackage{amstext}
\usepackage{epsfig}
\usepackage{latexsym}
\usepackage{amstext}
\usepackage{ulem}
\usepackage{amssymb}
\usepackage{array}
\usepackage{multirow}
\usepackage{tikz}
\usepackage{authblk}
\usepackage{cite}

\usepackage{amssymb}
\usepackage{array}
\usepackage{multirow}
\usepackage{tikz}
\newcommand{\ds}{\displaystyle}
\newcommand{\beq}{\begin{eqnarray}}
\newcommand{\eeq}{\end{eqnarray}}
\newcommand{\beqq}{\begin{eqnarray*}}
\newcommand{\eeqq}{\end{eqnarray*}}

\newcommand{\B}{\mbox{\boldmath$B$}}

\newcommand{\s}{\mbox{\boldmath$s$}}

\font\bb=msbm10 at 12pt

\def\eE{\hbox{\bb E}}

\begin{document}
\begin{center}
{\large \textbf{{Exit versus escape in a stochastic dynamical system of neuronal networks explains heterogenous bursting intervals}}}\\[5mm]
Lou Zonca\footnote{Sorbonne University, Pierre et Marie Curie Campus, 5 place Jussieu 75005 Paris, France.}$^{,2}$
David Holcman\footnote{Group of Applied Mathematics and Computational Biology, \'Ecole Normale Sup\'erieure, France.}
\end{center}
\date{}
\begin{abstract}
Neuronal networks can generate burst events. It remains unclear how to analyse interburst periods and their statistics. We study here the phase-space of a mean-field model, based on synaptic short-term changes, that exhibit burst and interburst dynamics and we identify that interburst corresponds to the escape from a basin of attraction. Using stochastic simulations, we report here that the distribution of the these durations do not match with the time to reach the boundary. We further analyse this phenomenon by studying a generic class of two-dimensional dynamical systems perturbed by small noise that exhibits two peculiar behaviors: 1- the maximum associated to the probability density function is not located at the point attractor, which came as a surprise. The distance between the maximum and the attractor increases with the noise amplitude $\sigma$, as we show using WKB approximation and numerical simulations. 2- For such systems, exiting from the basin of attraction is not sufficient to characterize the entire escape time, due to trajectories that can return several times inside the basin of attraction after crossing the boundary, before eventually escaping far away. To conclude, long-interburst durations are inherent properties of the dynamics and sould be expected in empirical time series.
\end{abstract}
\section*{Keywords}
Stochastic dynamical systems, Modeling neuronal networks, Asymptotic analysis, WKB, Eikonal equations, Escape from an attractor.
\section*{AMS classification}
37A50, 60G40, 92B25,41A60
\section{Introduction}
Trajectories of a dynamical system perturbed by small noise can escape from a basin of attraction and in general the dynamics presents large fluctuations away from a stable attractor \cite{dykman1994observable,dykman1994large,maier1993}. These perturbations can even induce switching in multi-stable systems. Noise can also enhance the response to periodic external stimuli, a phenomenon known as stochastic resonance \cite{lindner2004}. In the case of interaction between noise and a dynamics presenting a Hopf bifurcation, oscillations that would disappear in the deterministic case can be maintained. Finally, noise can induce a shift in bifurcation values \cite{schimanskyGeier1985} or can stabilize an unstable equilibrium \cite{arnold1979,arnold1990,Wihstutz_book2003}.\\
In the context of modeling biological neuronal networks, noise also plays a critical role in defining collective rhythms or large synchronization. To reduce the complexity and difficulty inherent in analyzing large neuronal ensembles, mean-field models are used to study averaged behavior, which corresponds to projecting a high dimensional system into a low dimension, as it is the case for modeling fast synaptic adaptations \cite{Tsodyks1997}. Such models are used to study bursting activity, synchronization and oscillations in excitatory neuronal networks \cite{BrunelHakim1999,Hansel2001,Holcman_Tsodyks2006,Barak2007,daoduc2015}. However, in such models the distribution of interburst durations remains unclear, although these durations have recently been shown to control to the overall neuronal networks dynamics \cite{Rouach_CxKO} and can even influence the bursting activity during epilepsy.\\
Stuyding the escape rate from a basin of attraction for a noisy dynamical system usually consists in collecting trajectories that terminate when they hit for the first time the boundary of the basin of attraction, which occurs with probability one \cite{Matkowsky1977exit, Schuss:Book}. The escape rate and the distribution of exit points can be computed in the small noise limit using WKB approximation. Another interesting property is that the exit point distribution peaks at a distance $O(\sqrt{\sigma})$ from the saddle-point (where $\sigma$ is the noise amplitude) \cite{Schuss1980,BobrovskySchuss1982}. Metrics relation can also play a role in shaping the dynamics, so that when a focus attractor falls into the boundary layer of the basin of attraction, escaping trajectories exhibit periodic oscillations leading to an escape time distribution which is not exponential, because several eigenvalues are necessary to describe the distribution \cite{maier1996,verechtchaguina2006_1,verechtchaguina2006_2,verechtchaguina2007,daoduc2016,daoducPRE}. \\ In the case of periodically-driven systems, the escape rate scales by the field intensity \cite{smelyanskiy1999time,dykman2005}. In all these examples, escape ends when a trajectory hits the separatrix for the first time which will not be the case for the systems we wish to study here. We consider here a class of dynamical systems perturbed by a white noise of small amplitude for which trajectories exiting the basin of attraction can reenter multiple times before eventually escaping to infinity. This effect requires to clarify the difference between exiting versus escaping that we explain below.\\
The manuscript is organized as follows: in the first part, we introduce a reduced stochastic dynamical system for bursting, based on modeling synaptic depression-facilitation for an excitatory neuronal network \cite{Tsodyks1997,daoduc2015}. The distribution of interburst intervals corresponds to the escape from an attractor and numerical simulations reveal a shift in the distribution of exit points and multiple returns inside the attractor. In the second part, we describe a generic two-dimensional dynamical system, containing an attractor and one saddle-point. We analyze the stochastic perturbation and show that the maximum of the probability density function of trajectories before escape, is not centered at the attractor, but at a shifted location that depends on the noise amplitude $\sigma$. Finally, we focus on the escape from the basin of attraction. After exiting, trajectories can return inside the basin of attraction multiple times before eventually escaping to infinity. To conclude, each excursion outside and inside the basin of attraction contributes to increase the total escape time by a factor between 2 to 3 compared to the first exit time, providing a novel explanation for the large tail distribution of interbust intervals in experimental time series \cite{Rouach_CxKO}.
\section{Modeling the interburst durations in neuronal networks}
\subsection{Noisy Depression-facilitation model}
Biological neuronal networks exhibit complex patterns of activity as revealed by time series of a single neuron \cite{hille1978,yuste2010}, a population or an entire Brain area \cite{niedermeyer2005}.  To analyse neuronal networks, mean-field models are used to formulate the dynamics as stochastic differential equations. \\
Bursting dynamics is a transient period of time where an ensemble of neurons discharge and these events can be accounted for by using short-term plasticity properties of synapses, such as the classical depression and facilitation \cite{Tsodyks1997,Barak2007,daoduc2015}. Bursting is followed by an interburst period, where the network is almost silent. Neuronal population bursts separated by long interbursts can result from two-state synaptic depression \cite{Guerrier2015}. However, the refractory period could also be the result of other mechanisms such as afterhyperpolarization (AHP), mediated by leading long-lasting voltage hyperpolarisation transient and generated by potassium channels \cite{AHPmodel}.\\
We focus here on a depression-facilitation short-term synaptic plasticity model of network neuronal bursting \cite{Tsodyks1997,daoduc2015frontiers,Holcman_Tsodyks2006}, which consists of three equations \eqref{sys} for the mean voltage $h$, the depression $y$, and the facilitation $x$. We recall that synaptic depression describes the possible depletion of vesicular pools, necessary for neurotransmission following an action potential. In this phenomenology,  facilitation is a synaptic mechanism that reflects a transient increase of the release vesicular probability, mediated possibly by an increase of the local calcium concentration in the pre-synaptic terminal. The associated equation is driven by two opposite forces: one is the return to an equilibrium $X$ with a time constant $t_f$ and the other is an increase induced by a mean firing rate $h^+=max(h,0)$. Similarly, the depression variable $y$ returns exponentially to steady state with a time constant $t_r$. It can also decrease following a firing rate $h^+$, proportional to the available fraction $y$ of vesicles and the facilitation $x$. The three coupled equations for the mean voltage $h$, the depression $y$, and the synaptic facilitation $x$ are
\beq \label{sys}
\tau \dot{h} &=& - h + Jxy h^+ +\sqrt{\tau}\sigma \dot{\omega}\nonumber\\
\dot{x} &=& \dfrac{X-x}{t_f} + K(1-x) h^+ \\
\dot{y} &=& \dfrac{1-y}{t_r} - L xy h^+ , \nonumber
\eeq
where the population average firing rate $h^+ = \max(h,0)$ is a linear threshold function of the synaptic current. The mean number of connections (synapses) per neuron is accounted for by the parameter $J$ and the term $Jxy$ reflects the combined effect of the short-term synaptic plasticity on the network activity. We previously distinguished \cite{daoduc2015} the parameters $K$ and $L$ which describe how the firing rate is transformed into molecular events that are changing the duration and probability of vesicular release. The time scales $t_f$ and $t_r$ define the recovery of a synapse from the network activity. Finally, $\dot \omega$ is an additive Gaussian noise and $\sigma$ its amplitude.  We shall focus now on a reduced version of this system to study the distribution of interburst intervals which are interpreted as escape from a basin of attraction that we will define below, with similar properties as the ones presented in equation \eqref{Drift}.
\subsection{Reduction to a two-dimensional system and phase-space analysis}
System \eqref{sys} has 3 critical points: one attractor and two saddle-points. At the attractor $A=(0,X,1)$, the dynamics is very anisotropic \\
$\left(|\lambda_1|= \cfrac{1-JX}{\tau} \approx 12.6 \gg |\lambda_2| = \cfrac{1}{\tau_f} \approx 1.1 \gg |\lambda_3| \approx \cfrac{1}{\tau_r} = 0.34, \text{using parameters in table \ref{tableParam}}\right)$ and thus can be reduced to a 2D-plan $y=Cte$ so that
\beq
\dot{y}=0=\dfrac{1-y}{\tau_r} - L xy h^+ = 0 \iff y = \cfrac{1}{1+\tau_r Lxh^+},
\eeq
and we obtain the simplified system:
\beq\label{2Dsyst}
\begin{array}{r c l}
	\arraycolsep=1.4pt\def\arraystretch{2.5}
	\dot{h}&=&\cfrac{h\left(Jx-1-\tau_rLxh^+ \right)}{\tau(1+\tau_rLxh^+)}+\sqrt{\tau}\sigma \dot{\omega}\\
	\dot{x} &=& \cfrac{X-x}{\tau_f} + K(1-x) h^+.
\end{array}
\eeq
\begin{figure}[http!]
\centering
	\includegraphics[scale=0.74]{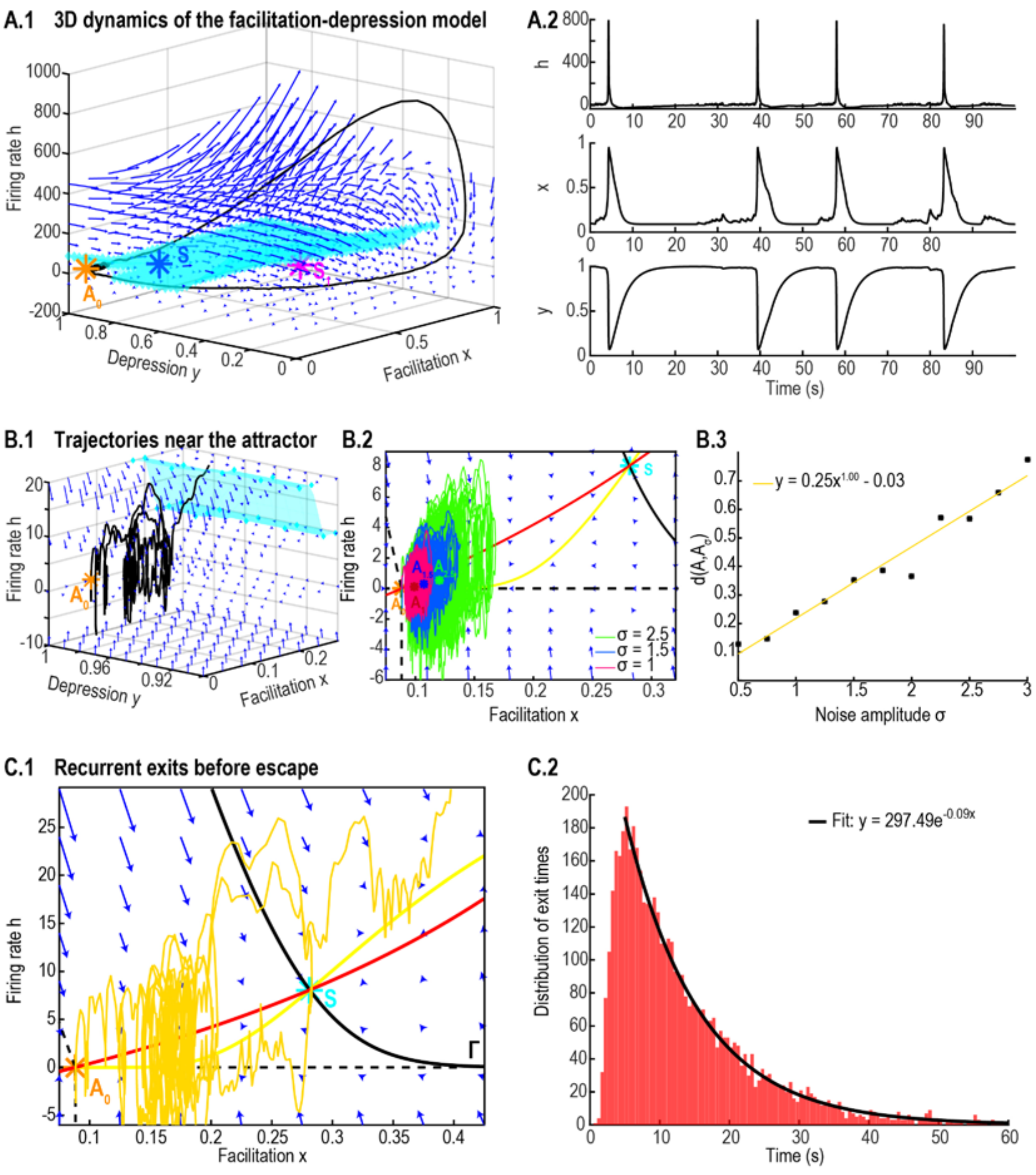}
	\caption{ \textbf{Modeling Bursting and interbursting using a facilitation-depression model} \textbf{A. Dynamics generated by system \eqref{sys}. A.1} a bursting trajectory (black) in the phase-space, characterized by an attractor $A_0$ (yellow), two saddle-points $S$ (blue) and $S_1$ (pink) with a separatrix delimiting the basin of attraction of $A_0$ (cyan surface). \textbf{A.2} Simulated time-series of system \eqref{sys} with a noise amplitude $\sigma=5$, mean voltage $h$ (upper), facilitation $x$ (center) and depression $y$ (lower) for $T=100s$. \textbf{B. Dynamics in the basin of attraction of $A_0$. B.1} Inset near $A_0$ showing an escaping trajectory (black). \textbf{B.2} Phase-space of the 2D projected dynamics \eqref{2Dsyst} for several trajectories ($T=500s$) associated with three noise levels $\sigma = 1$ (pink),1.5 (blue) and 2.5 (green) rotating around shifted attractors $A_\sigma$ (centers of mass of the trajectories). \textbf{B.3} Distance d($A_0$,$A_\sigma$) for $\sigma \in [0.5,3]$ compared to a numerical fit (yellow). \textbf{C. Escape dynamics with several returns inside the basin of attraction. C.1} Trajectory exiting the basin of attraction. \textbf{C.2} Distribution of exit times where the tail is well approximated by a single exponential with $\lambda_e=0.09$ (or $\bar{\tau}_e~11s$).} \label{application}
\end{figure}
The deterministic system \eqref{2Dsyst} for $\sigma=0$ has 3 critical points, two attractors and one saddle-point:
{\bf Attractor $A_0$ }is given by $h=0$ and $x=X$. The Jacobian is
\beq\label{jac_A}
J_{A_0} = \arraycolsep=1.4pt\def\arraystretch{2.0}
\left(
\begin{array}{c c}
	\cfrac{-1+JX}{\tau} \phantom{1234}& 0 \\
	K(1-X) \phantom{1234}& - \cfrac{1}{\tau_f}\\
\end{array}
\right).
\eeq
With the  parameters defined in table \ref{tableParam}, the eigenvalues are $(\lambda_1,\lambda_2)  = \left(\cfrac{JX-1}{\tau}, \cfrac{1}{\tau_f}\right) \approx (-12.6, -1.11)$.\\
{\bf Saddle-point $S$} has coordinates $S_1 (h_1 \approx 8.07; x_1 \approx 0.28)$. Its eigenvalues are  $(\lambda_1,\lambda_2) \approx  (-5.73,1.43)$ .\\
{\bf Attractor $A_2$} is given by  $A_2 (h_2 \approx 28.8; x_2 \approx 0.53)$. Its eigenvalues are $(\lambda_1,\lambda_2) \approx (-11.9,-1.33)$.\\
The two attractors are separated by a 1D stable manifold $\Gamma$ passing through the saddle-point $S$ (fig. \ref{application}A, solid black). To study the dynamics around the attractor $A_0$, where we approximate $\dot{y}=0$,  we first generated stochastic trajectories and observe two novel phenomena: 1) in the basin of attraction before exiting, trajectories fluctuate around a point not centered at the attractor $A_0$, but rather around a shifted point $A_\sigma$ the position of which depends on the noise amplitude (fig. \ref{application}B.2-B.3); 2) trajectories that exit the basin of attraction through the separatrix $\Gamma$ can reenter multiple times before finally escaping. In the reduced system \ref{2Dsyst}, escape is characterized by falling to the second attractor $A_2$.  The most interesting unexplain phenomena revealed by figs. \ref{application}C.1-C.2 is the single exponential decay rate with a decay $\lambda=0.09$ (or $\bar{\tau}_e=11s$). This is in contrast with the mean escape time $\langle\tau_0\rangle=4.35$ (estimated numerically) from the attractor for a trajectory starting at the attractor $A$ and reaching the separatrix $\Gamma$. The rest of the manuscript is dedicated to understand how this discrepancy can be resolved and also to better study the properties 1) and 2) we identified numerically. For that purpose, we study below a generic dynamical system that serves as a model and obtain specific computational criteria and finally, we resolve the present enigma.
\section{When a perturbation of a two-dimensional system by a Gaussian noise induces a shift of the density function peak with respect to the attractor position} \label{2D_description}
\begin{figure}[http!]
	\includegraphics[scale=0.74]{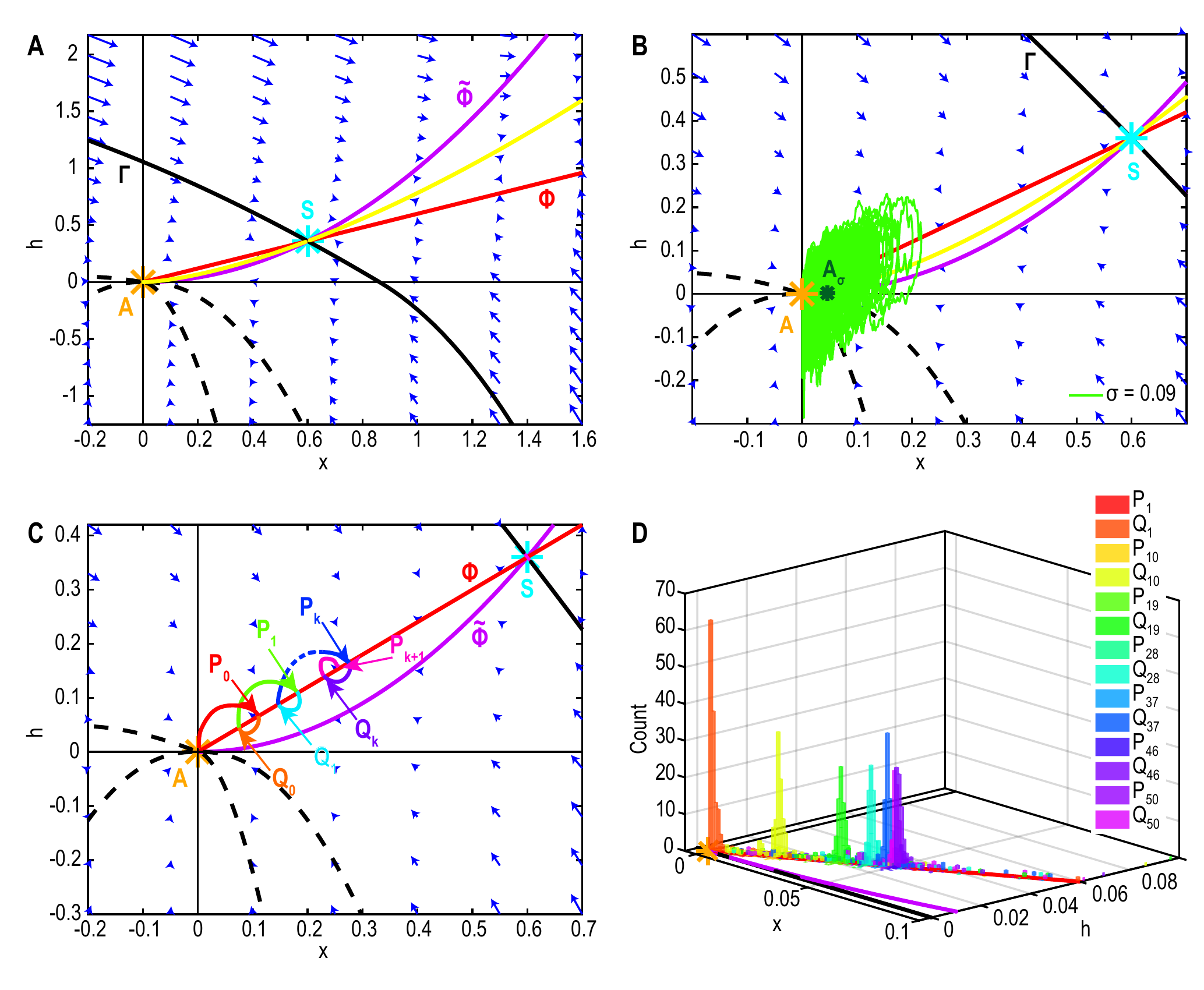}
	\caption{\textbf{Emergence of a shift in the attractor's position for the noisy dynamical system \eqref{Drift}.} \textbf{A.} Phase-space of system \eqref{Drift}. The basin of attraction associated to the attractor $A$ (yellow) has two stable manifolds (dashed lines) and is delimited by the stable manifold $\Gamma$ (solid black) passing through the saddle-point $S$ (cyan). \textbf{B.} Stochastic trajectory (green) for a noise amplitude $\sigma = 0.09$. The center of mass $A_\sigma$ is shifted towards the right with $x_{A_\sigma}\approx 0.05$. \textbf{C.} Successive intersection points $P_k$ and $Q_k$ of trajectories with $\Phi_x$ (red). \textbf{D.} Distributions $\rho_{P_k}$ and $\rho_{Q_k}$ (500 runs, $\sigma=0.09$) for $1\leq k\leq 50$. The peak of the converging distributions $\rho_{P_{46}}$ to $\rho_{P_{50}}$ (purple) indicates the x-coordinate of the shifted attractor $x_{A_\sigma}\approx 0.05$.}\label{PhaseSpace}
\end{figure}

We consider a class of two-dimensional stochastic dynamical system described by
\beq\label{Drift}
\begin{array}{r c l}
	\arraycolsep=1.4pt\def\arraystretch{2.5}
	\dot{h}&=&-\alpha h + x^2 +\sigma \dot{\omega}=b_1(\s)\\
	\dot{x} &=& F(h,x)=b_2(\s),
\end{array}
\eeq
where
\beq
F(h,x)=\left\{
\begin{array}{l c l}
	\arraycolsep=1.4pt\def\arraystretch{2.5}
	h - \gamma x &\text{for}& h\geq 0\\
	- \gamma x &\text{for}& h\leq 0,\\
 \end{array} \right.
\eeq
We rewrite this process with $\s=(x,h)$
\beq \label{stocProc}
d\s = \B(\s)dt + \Xi dW,
\eeq
where
\beq \label{B}
\B(\s) = \begin{pmatrix}
b_1(\s) \\
b_2(\s)
\end{pmatrix} \hbox{ and }
\Xi= \begin{pmatrix}
	\sqrt{\sigma} & 0 \\
	0 & 0
\end{pmatrix}.
\eeq
In the following, $\alpha \in ]0,1]$, $\gamma \in ]0, \alpha[$, $\dot{\omega}$ is a Gaussian white noise and $\sigma$ its amplitude. Our goal here is to study some properties of such systems. This system has two critical points, $A = (0,0)$ (fig. \ref{PhaseSpace}A yellow star) and $S = (\gamma^2 \alpha, \gamma \alpha)$ (fig. \ref{PhaseSpace}A cyan star). The jacobian of the system at point $A$ can be computed either for $h \geq 0$ or for $h \leq 0$ and in both cases, we have
\beq\label{jacAnoDrift}
J_A =
\begin{pmatrix}
	-\alpha & 0 \\
	1 & -\gamma\\
\end{pmatrix}.
\eeq
The attractor $A$ has real eigenvalues $\lambda_1 = -\alpha$ and $\lambda_2 = -\gamma$ (its stable manifolds are shown in fig. \ref{PhaseSpace}A, dotted black lines). The first coordinate of the point $S$ is $h_S=\gamma^2\alpha>0$ and the jacobian is
\beq\label{jacSnoDrift}
J_S =
\begin{pmatrix}
	-\alpha & 2\alpha \gamma \\
	1 & -\gamma\\
\end{pmatrix}.
\eeq
Both eigenvalues are real, $\lambda_\pm = -\cfrac{1}{2}\left(-(\alpha+\gamma)\pm \sqrt{(\alpha+\gamma)^2+4\alpha \gamma}\right)$ and thus $S$ is a saddle point (with $\alpha = 1$ and $\gamma = 0.6$ we have  $\lambda_+ \approx 0.314$ and $\lambda_-\approx -1.914$). \\
The separatrix that delimits the basin of attraction of $A$ is the stable manifold of $S$ (fig. \ref{PhaseSpace}A solid black curve). As we shall describe below the unstable manifold defines the escaping direction (fig. \ref{PhaseSpace}A yellow curve). It is located between the $x$ (respectively $h$) nullcline $\Phi_x=\{(x,h) | h=\gamma x\}$, \ref{PhaseSpace}A red (respectively $\tilde{\Phi}_h=\{(x,h) |h=x^2/\alpha\}$, purple).\\
Numerical simulations reveal that the stochastic trajectories are not centered around the deterministic attractor $A$ (fig. \ref{PhaseSpace}B, green trajectory). Indeed, we computed the shifted attractor as the expectation of the center of mass for each trajectory, before escaping at time $\tau_{\omega}$,
\beq
A_\sigma=\eE_{\omega}[\frac{1}{\tau_{\omega}}\int_{0}^{\tau_{\omega}} \s_{\omega}(t)dt].
\eeq
The empirical distribution peaks at the point $A_\sigma$, which we found to be shifted towards the right of the attractor $A$ (fig. \ref{PhaseSpace}B, green star). Our next goal is to study how this shift depends on $\sigma$.
\subsection{Numerical study of a shifted maximum density at \boldmath $A_\sigma$ \unboldmath}
To better characterize the shifted peak $A_\sigma$  induced by the noise for the maximum of the density position, before trajectories escape the attractor, we ran simulations of model eq. \eqref{Drift} (fig. \ref{PhaseSpace}B $\sigma =0.09$), where trajectories are simulated for 500s. We observed that trajectories are looping around the $x$-nullcline $\Phi_x$ (red). To characterize the shift in the distribution, we study the distribution of points $P_0$
\beq
\rho_{P_0} = P(\s(t)\in \Phi | \s(0)=A)
\eeq
where a trajectory starting at $A$ hits $\Phi$ for the first time (fig. \ref{PhaseSpace}C, red). We generated the empirical distribution $\rho_{P_0}$ by simulating 150 trajectories (fig. \ref{PhaseSpace}D, red) and we found that this distribution is peaked at $P_0$ close to $A$. To further understand the dynamics we then investigated the distribution of points $Q_0$
\beq
\rho_{Q_0}=P(\s(t)\in \Phi | \s(0)=P_0)
\eeq
where a trajectory starting at $P_0$ hits $\Phi_x$ for the first time (fig. \ref{PhaseSpace}C, orange). This distribution is also peaked and located nearby $\rho_{P_0}$ (fig. \ref{PhaseSpace}D, orange). We then iterated this process to obtain the successive distributions $P_k$
\beq
\rho_{P_k}=P(\s(t)\in \Phi | \s(0)=Q_{k-1}),
\eeq
of points where a trajectory starting initially at $Q_{k-1}$ hits $\Phi_x$. Similarly, we define the distributions of the points
\beq
\rho_{Q_k}=P(\s(t)\in \Phi | \s(0)=P_k)
\eeq
where a trajectory starting at the peak of $\rho_{P_k}$ hits $\Phi_x$ for the first time (fig. \ref{PhaseSpace}D). Interestingly, we observed that the distributions are peaked and progress along $\Phi_x$ towards the separatrix $\Gamma$. However, after a few iterations, the successive distributions $\rho_{P_k}$ and $\rho_{Q_k}$, seems to accumulate toward a shifted equilibrium (fig. \ref{PhaseSpace}D, pink and purple distributions).
\subsection{Computing the steady-state distribution and the distance of the maximum to the attractor \boldmath $A$ \unboldmath using WKB approximation}\label{pseudoEqLoc}
\subsubsection{Steady state Fokker-Planck equation for system \eqref{Drift}}\label{pseudoEqLocPDF}
In this subsection, we study the probability density function (pdf) of the system \eqref{Drift} for trajectories that stay inside the basin of attraction of $A$. We first generated 300s simulations for three values of the noise amplitude $\sigma=0.03$ (pink), 0.09 (blue) and 0.12 green (fig. \ref{phaseSpace_PDF}A) showing that the pseudo-distributions for points that do not escape the basin of attraction are peaked at $A_\sigma$ shifted to the right from $A$. We now compute this pdf using WKB approximation \cite{Schuss1980}. The steady-state pdf $p$ satisfies the stationary Fokker-Planck Equation (FPE)
\beq\label{FPE}
\cfrac{\sigma}{2}\cfrac{\partial^2 p}{\partial h^2} -(\nabla \cdot \B) p - \B \cdot \nabla p = -\delta_A,
\eeq
where $\delta_A$ is the $\delta$-Dirac function at point $A$. Due to the discontinuity of the field at $h=0$ we compute $\nabla \cdot \B$ on the two half spaces $(h \geq 0)$ and $(h \leq 0)$ separately. In the small noise limit $\sigma \to 0$, the WKB solution has the form
\beq\label{WKB}
p(\s) = K_{\sigma}(\s) e^{-\cfrac{\psi(\s)}{\sigma}},
\eeq
where $K_{\sigma}$ is a regular function that admits an expansion
\beq\label{K}
K_{\sigma}(\s)=\sum_{i=0}^{\infty} K_i(\s) \sigma^i.
\eeq
The eikonal equation is obtained by injecting \eqref{WKB} in \eqref{FPE} and by keeping only the higher order terms in $\sigma$ (ie $\sigma^{-1}$)
\beq\label{eikonal}
\B \cdot \nabla \psi + \cfrac{1}{2} \left(\cfrac{\partial \psi}{\partial h}\right)^2 = 0
\eeq
and the transport equation is obtained using the order 1 terms:
\beq\label{transport}
\B \cdot \nabla K_0 + \cfrac{\partial \psi}{\partial h}\cfrac{\partial K_0}{\partial h}=-\left(\nabla \cdot \B + \cfrac{1}{2}\cfrac{\partial ^2 \psi}{\partial h^2}\right) K_0.
\eeq
To solve \eqref{eikonal}, we use the method of characteristics with notation  $q=(q_1,q_2)=\nabla \psi$. Then eq. \eqref{eikonal} becomes
\beq
F(\s,q,\psi)=\B \cdot q + \cfrac{1}{2}q_1^2=0
\eeq
and the characteristics are given by
\beq\label{characteristics}
\arraycolsep=1.4pt\def\arraystretch{2.5}
\begin{array}{r c l}
	\cfrac{dh}{dt}&=&b_1(\s)+q_1\\
	\cfrac{dx}{dt}&=&b_2(\s)\\
	\cfrac{dq_1}{dt}&=&-F_h=-\cfrac{\partial b_1}{\partial h}q_1-\cfrac{\partial b_2}{\partial h}q_2\\
	\cfrac{dq_2}{dt}&=&-F_x=-\cfrac{\partial b_1}{\partial x}q_1-\cfrac{\partial b_2}{\partial x}q_2\\
	\cfrac{d\psi}{dt}&=&\cfrac{1}{2} q_1^2.\\
\end{array}\
\eeq
To define the initial condition, we can choose a neighborhood $V_A$ of $A$ (positioned at the origin), where $\psi$ has a quadratic approximation
\beq
\psi(\s) \approx \cfrac{1}{2}\s^TR\s +o(|s|^2)) \hbox{ for } \s \in V_A,
\eeq
and $R$ is a symmetric positive definite matrix defined by a degenerated matrix equation at $A$
\beq
(J_A \s)^T \cdot \nabla \psi + \cfrac{1}{2}q_1^2 = 0,
\eeq
where $J_A$ is the Jacobian matrix at point $A$ defined by relation \eqref{jacAnoDrift}. We obtain
\beq \label{psiApprox}
\psi(\s) \approx \cfrac{1}{2}\s^T\begin{pmatrix}
	2\alpha & 2\gamma\\
	2\gamma & 2\gamma
\end{pmatrix} \s.
\eeq
The $\psi$ contours are the ellipsoids given by
\beq\label{ellipsoids}
\alpha h^2+ 2\gamma xh + \gamma x^2=\epsilon,
\eeq
for small $\epsilon>0$. To conclude, we choose for the initial conditions one of the small ellipsoids given by \eqref{ellipsoids} by fixing later on the value of $\epsilon$.
\subsubsection{Solution in the subspace \boldmath $ h \leq 0 $ \unboldmath}
A direct integration of system \eqref{characteristics} gives for $t\geq0$
\beq \label{characHneg}
\arraycolsep=1.4pt\def\arraystretch{2.5}
\begin{array}{r c l}
	h(t)&=&\left(h_0-\cfrac{ x_0^2}{\alpha-2\gamma}-\cfrac{q_{1,0}}{2\alpha}\right)e^{\ds -\alpha t} + \cfrac{x_0^2}{\alpha -2\gamma}e^{\ds -2\gamma t} +\cfrac{q_{1,0}}{2\alpha}e^{\ds \alpha t}\\
	x(t)&=&x_0e^{\ds -\gamma t}\\
	q_1(t)&=&q_{1,0}e^{\ds \alpha t} \\
	q_2(t)&=&\left(q_{2,0}-\cfrac{2 x_0 q_{1,0}}{\alpha-2\gamma}\right)e^{\ds \gamma t}-\cfrac{2 x_0 q_{1,0}}{\alpha-2\gamma}e^{\ds (-\gamma+\alpha)t}\\
	\psi(t)&=&\cfrac{q_{1,0}^2}{4 \alpha}e^{\ds 2\alpha t},
\end{array}
\eeq
where the initial conditions are $h(0)=h_0, x(0)=x_0, q_1(0)=q_{1,0} \text{ and } q_2(0)=q_{2,0}$. Substituting the expression of $x$ and $q_1$ in $h$, we obtain
\beq \label{psi_xh_hNeg}
\psi(h,x)=\alpha\left(h-	\left(h_0-\cfrac{ x_0^2}{\alpha-2\gamma}-\cfrac{q_{1,0}}{2\alpha}\right) \left(\cfrac{x}{x_0}\right)^{\ds \cfrac{\alpha}{\gamma}} - \cfrac{x^2}{\alpha-2\gamma}\right)^2.
\eeq
We solve the transport equation \eqref{transport} along the characteristics \eqref{characHneg}. Using \eqref{Drift} and \eqref{psi_xh_hNeg}, we obtain
\beq
\cfrac{d K_0(\s(t))}{dt}=\gamma K_0(\s(t)),
\eeq
yielding
\beq \label{K0hNeg}
K_0(s(t))=C e^{\ds \gamma t}.
\eeq
For $\alpha > \gamma$, and choosing $\s(t)\in V_A,$
\beq \label{absCurv}
\tilde{\s}(t) \approx \int_0^t x_0 e^{-\gamma u} du \approx - x_0 \cfrac{e^{\ds -\gamma t}-1}{\gamma}.
\eeq
Thus
\beq
K_0 \sim \cfrac{x_0}{x_0 + \gamma \s \cdot e_2} =  \cfrac{x_0}{x_0 + \gamma x},
\eeq
where $e_2=(0,1)^T$ is the eigenvector associated to the eigenvalue $\lambda_2$=$-\gamma$. Finally,
\beq \label{pdfFinal_hneg}
p(\s) \sim \cfrac{x_0}{x_0 + \gamma x} e^{-\cfrac{\psi(\s)}{\sigma}}.
\eeq
\subsubsection{Solution in the subspace \boldmath $ h \geq 0 $ \unboldmath}
\begin{figure}[http!]
\centering
\includegraphics[scale=0.74]{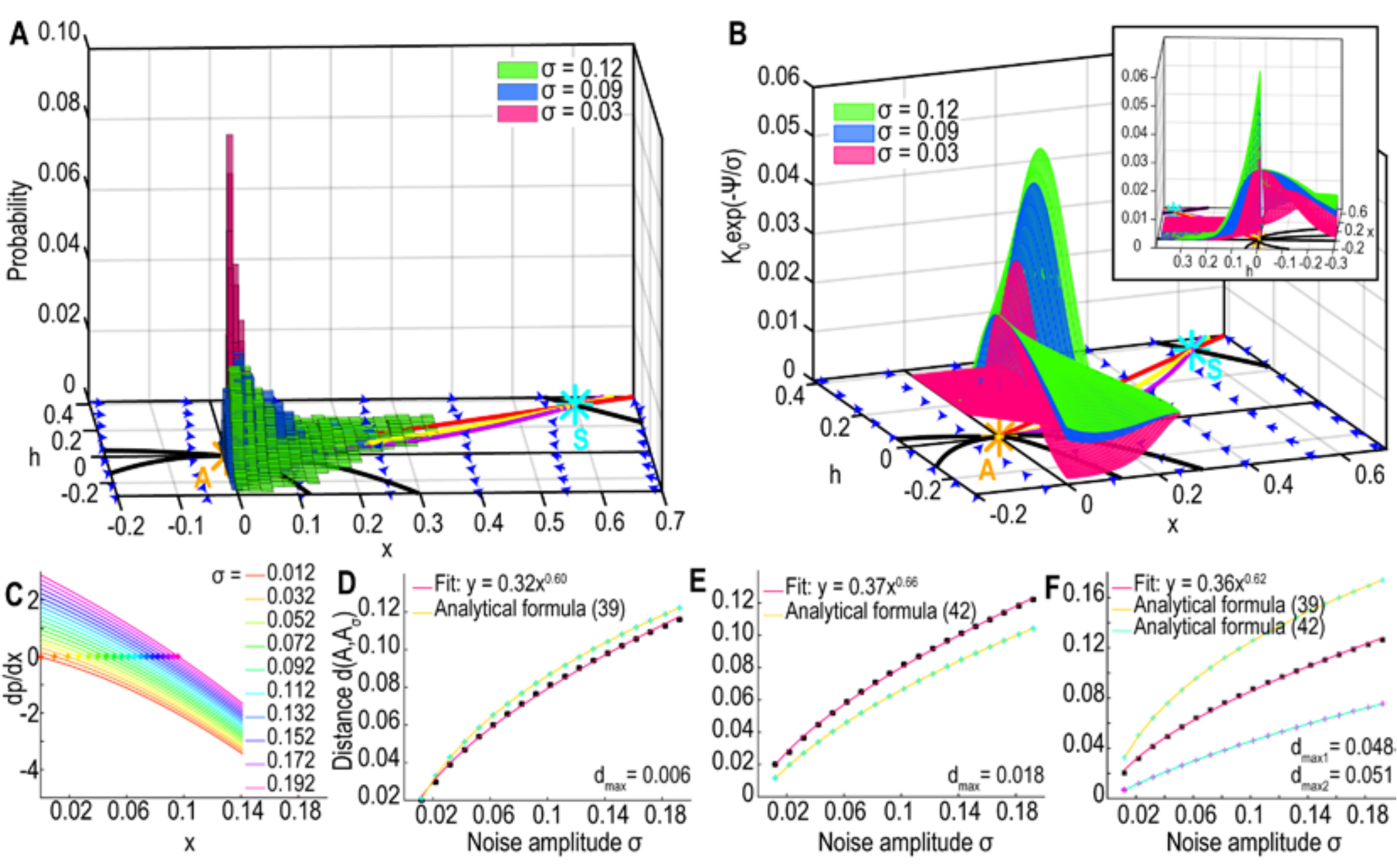}
\caption{\textbf{Position of the shifted attractor \boldmath $A_\sigma$ \unboldmath }\textbf{A.} Simulated pdf of the trajectories for three noise levels $\sigma = 0.03$ (pink), $\sigma = 0.09$ (blue) and $\sigma = 0.12$ (green) for $\gamma = 0.6$ and $\alpha = 1$. \textbf{B.} Analytical distributions for the three noise levels. Inset: same distributions with a different perspective. \textbf{C.} $\cfrac{\partial p}{\partial x}_{|h=0^+}(x)$ for $\sigma \in [0.01,0.2]$ (the crosses indicate the zeros). \textbf{D-E.} Distance $d(A,A_\sigma)$ as a function of $\sigma$. Numerical solution (black stars with numerical fit in pink) and analytical relation \ref{xM_06} (resp. \ref{xM_09}, cyan crosses with yellow curve) for $\gamma=0.6$ (resp. 0.9) and $\alpha = 1$. \textbf{F.} Distance $d(A,A_\sigma)$ as a function of $\sigma$. Numerical solution (black stars with numerical fit in pink) compared with the analytical expressions \eqref{xM_06} and \eqref{xM_09} (cyan crosses with yellow curve and magenta crosses with blue curve) in the case $\gamma=0.75$ and $\alpha = 1$ for which neither polynomial approximation is valid.}\label{phaseSpace_PDF}
\end{figure}

In this case, we cannot integrate system \eqref{characteristics} analytically. In the neighborhood of $A$, $x\ll 1$ and since $\psi$ is a smooth function, we can neglect the quadratic terms in system \eqref{characteristics} yielding
\beq \label{linCharacHpos}
\arraycolsep=1.4pt\def\arraystretch{2.5}
\begin{array}{r c l}
	\cfrac{dh}{dt}&\approx&-\alpha h+q_1\\
	\cfrac{dx}{dt}&=&-\gamma x + h\\
	\cfrac{dq_1}{dt}&=&\alpha q_1-q_2\\
	\cfrac{dq_2}{dt}&\approx&-F_x= \gamma q_2\\
	\cfrac{d\psi}{dt}&=&\cfrac{1}{2} q_1^2.\\
\end{array}\
\eeq
Integration of system \eqref{linCharacHpos} gives, for $t\geq 0$
\beq \label{characHpos}
\arraycolsep=1.4pt\def\arraystretch{2.5}
\begin{array}{r c l}
	h(t)&=&H_0e^{\ds -\alpha t} + \cfrac{Q_0}{2\alpha}e^{\ds \alpha t} - \cfrac{q_{2,0}}{\gamma^2-\alpha^2}e^{\ds \gamma t}\\
	
	x(t)&=&X_0e^{\ds -\gamma t}+\cfrac{H_0}{\gamma-\alpha}e^{\ds -\alpha t} +\cfrac{Q_0}{2\alpha (\gamma+\alpha)}e^{\ds \alpha t} - \cfrac{q_{2,0}}{2\gamma(\gamma^2-\alpha^2)}e^{\ds \gamma t}\\
	
	q_1(t)&=&Q_0e^{\ds \alpha t} -\cfrac{q_{2,0}}{\gamma -\alpha}e^{\ds \gamma t}\\
	
	q_2(t)&=&q_{2,0}e^{\ds \gamma t}\\
	
	\psi(t)&=&\cfrac{Q_0^2}{4\alpha}e^{\ds 2\alpha t}+\cfrac{q_{2,0}^2}{4\gamma(\gamma -\alpha)^2}e^{\ds 2 \gamma t}-\cfrac{Q_0 q_{2,0}}{\gamma^2-\alpha^2}e^{\ds (\gamma+\alpha)t},
\end{array}
\eeq
where
\beq
\arraycolsep=1.4pt\def\arraystretch{2.5}
\begin{array}{r c l}
	Q_0&=&q_{1,0}+\cfrac{q_{2,0}}{\gamma-\alpha}\\
	H_0&=&h_0-\cfrac{Q_0}{2\alpha}+\cfrac{q_{2,0}}{\gamma^2-\alpha^2}\\
	X_0&=&x_0-\cfrac{H_0}{\gamma-\alpha}-\cfrac{Q_0}{2\alpha(\gamma+\alpha)}+\cfrac{q_{2,0}}{2\gamma(\gamma^2-\alpha^2)}
\end{array}
\eeq
and  the initial conditions are $h(0)=h_0, x(0)=x_0, q_1(0)=q_{1,0} \text{ and } q_2(0)=q_{2,0}$.\\
To derive the eikonnal solution, we eliminate the time and start with the relation
\beq \label{approx}
q_2 \approx 2\gamma (h+x),
\eeq
which is obtained from \eqref{psiApprox}. Substituting \eqref{approx} in $h$, we obtain near the attractor $A$
\beq \label{psi_xh_hpos}
\psi(h,x) \approx \cfrac{Q_0^2}{4\alpha}\left(\cfrac{2\gamma (h+x)}{q_{2,0}}\right)^{\cfrac{2\alpha}{\gamma}}+\cfrac{\gamma(h+x)^2}{(\gamma-\alpha)^2}+\cfrac{Q_0 q_{2,0}}{\gamma^2-\alpha^2}\left(\cfrac{2\gamma(h+x)}{q_{2,0}}\right)^{\cfrac{\alpha+\gamma}{\gamma}}.
\eeq
Furthermore, we solve the transport equation \eqref{transport} along the characteristics \eqref{characHpos}. We differentiate twice \eqref{psi_xh_hpos} with respect to $h$, we obtain $\cfrac{\partial ^2 \psi}{\partial h^2} \approx \cfrac{\gamma}{(\alpha-\gamma)^2}$, which leads to
\beq
\cfrac{d K_0(\s(t))}{dt} \approx \left(\alpha+\gamma -\cfrac{\gamma}{(\gamma-\alpha)^2}\right)K_0(\s(t)).
\eeq
Using \eqref{absCurv} we obtain
\beq \label{K0_hpos}
K_0 \sim \left(\cfrac{x_0}{x_0 + \gamma x}\right)^{\cfrac{\alpha+\gamma}{\gamma}-\cfrac{1}{(\alpha-\gamma)^2}}.
\eeq
Finally, for $\s \in V_A$
\beq\label{pdfFinal}
p(\s)\sim K_0(\s) e^{-\cfrac{\psi(\s)}{\sigma}},
\eeq
where $K_0$ and $\psi$ are defined by relations \eqref{K0_hpos} and \eqref{psi_xh_hpos}, respectively. When $\alpha \neq \gamma$ and $\alpha \neq 2\gamma$, the exponent in \eqref{K0_hpos} is positive when
\beq
\cfrac{\alpha+\gamma}{\gamma} -\cfrac{1}{(\gamma-\alpha)^2}>0 \iff 1-\left(\cfrac{\gamma}{\alpha}\right)-\left(\cfrac{\gamma}{\alpha}\right)^2+\left(\cfrac{\gamma}{\alpha}\right)^3-\cfrac{\gamma}{\alpha^3}>0,
\eeq
that is for $\cfrac{\gamma}{\alpha}>0.45$. For the range of parameters $0.45\alpha<\gamma<\alpha$ and $\alpha \neq 2\gamma$, the pdf has a maximum located on the $h=0^+$ axis shifted towards the right of $A$ (fig. \ref{phaseSpace_PDF}B, for three values of the noise amplitude $\sigma=0.03$ (pink), 0.09 (blue) and 0.12 green and for $\gamma=0.6$). This maximum gives the position of the shifted attractor $A_\sigma$, which depends on the noise amplitude. When $\cfrac{\gamma}{\alpha}\leq0.45$, since the linearization approximation in \eqref{approx} is not valid, formula \eqref{pdfFinal} cannot be use to approximate the pdf.
\subsubsection{Computing the distance between \boldmath $A_\sigma$ \unboldmath and \boldmath $A$ \unboldmath}\label{pseudoEqLocSol}
To study how the distance between $A$ and $A_\sigma$ depends on the noise amplitude $\sigma$, we use that the maximum of the pdf \eqref{pdfFinal} is given by
\beq
\nabla p = 0.
\eeq
However, the partial derivative along $h$ is discontinuous, and the analytical expression for pdf $p$ (\ref{pdfFinal_hneg}, \ref{pdfFinal}) is decreasing with $|h|$ on both halves of the phase-space. Thus based on the numerical motivation that the maximum of the pdf $A_\sigma$ is shifted along the $x$ axis, we are left with solving $\cfrac{\partial p}{\partial x}_{|h=0^+}=0$. Using relation \ref{pdfFinal}, for $h \geq 0$ we obtain
\beq \label{gradPeq0}
\begin{split}
	-\cfrac{\left(\cfrac{\alpha + \gamma}{\gamma}-\cfrac{1}{(\alpha-\gamma)^2}\right) \gamma}{x_0+\gamma x}+\cfrac{1}{\sigma}\left(-\cfrac{Q_0^2}{q_{2,0}}\left(\cfrac{2\gamma}{q_{2,0}}\right)^{\ds \frac{2\alpha}{\gamma}-1}x^{\ds \frac{2\alpha}{\gamma}-1}\right.\\
	\left.-\cfrac{2Q_0}{\gamma-\alpha}\left(\cfrac{2\gamma}{q_{2,0}}\right)^{\ds \frac{\alpha}{\gamma}}x^{\ds \frac{\alpha}{\gamma}}
	-\cfrac{2\gamma x}{(\alpha-\gamma)^2}\right)=0,
\end{split}
\eeq
We rewrite \eqref{gradPeq0} as
\beq\label{dpdxEq0}
-\cfrac{A1}{1+\cfrac{\gamma}{x_0}x} - \cfrac{1}{\sigma}\left(A_2x^{\ds \frac{2\alpha}{\gamma}-1}+A_3x^{\ds \frac{\alpha}{\gamma}}+A_4x\right)=0,
\eeq
where
\beq
A_1 = \left(\cfrac{\alpha+\gamma}{\gamma}-\cfrac{1}{(\alpha-\gamma)^2}\right)\cfrac{\gamma}{x_0},
A_2 = \cfrac{Q_0^2}{q_{2,0}}\left(\cfrac{2\gamma}{q_{2,0}}\right)^{\ds \frac{2\alpha}{\gamma}-1},
A_3 = \cfrac{2Q_0}{\gamma-\alpha}\left(\cfrac{2\gamma}{q_{2,0}}\right)^{\ds \frac{\alpha}{\gamma}},
A_4 = \cfrac{2\gamma}{(\alpha-\gamma)^2}.
\eeq
The algebraic equation \ref{gradPeq0} describes the shift in the $x$ axis of the pdf peak compared to the attractor. This equation cannot be solved analytically in general and we solved it numerically for various values of $\sigma$ (fig. \ref{phaseSpace_PDF}C). \\
However, we shall describe two cases for which equation \eqref{dpdxEq0} can be approximated by a polynomial equation. In the range $0.5<\cfrac{\gamma}{\alpha}\leq0.645$ and $0.885\leq\cfrac{\gamma}{\alpha}<1$, we can compute the absolute difference $d(\sigma)=|x_{M,num}(\sigma)-x_M(\sigma)|$ between the numerical result $x_{M,num}(\sigma)$ and the solution of the approximated polynomial equation $x_M(\sigma)$ that we define below (equations \ref{xM_06} and \ref{xM_09}) for $\sigma \in [0,0.2]$ and by using the following criteria $d_{max}=\max\limits_{\sigma \in [0,0.2]}(d(\sigma))<0.02$ (fig. \ref{phaseSpace_PDF}D-F).
\subsubsection{Approximated expression for the distance \boldmath $A_\sigma-A$ \unboldmath in the range \boldmath $0.5<\cfrac{\gamma}{\alpha}<0.645$}
In the mentioned range, we approximate $\cfrac{2\alpha}{\gamma}-1 \approx 2$ and $\cfrac{\alpha}{\gamma} \approx 2$ and \eqref{dpdxEq0} becomes the third order polynomial equation
\beq\label{ploy3approx}
A_1\sigma + A_4x + \left(A_2+A_3+\cfrac{A_4\gamma}{x_0}\right)x^2+ (A_2+A_3)\cfrac{\gamma}{x_0}x^3 = 0.
\eeq
The solution is (fig. \ref{phaseSpace_PDF}D, cyan crosses and yellow curve)
\beq\label{xM_06}
x_M(\sigma) = \left(\cfrac{-q(\sigma)-\sqrt{\Delta(\sigma)}}{2}\right)^{\ds 1/3}+\left(\cfrac{-q(\sigma)+\sqrt{\Delta(\sigma)}}{2}\right)^{\ds 1/3}-\cfrac{c_2}{3c_1},
\eeq
where
\beq\label{interParam_06}
\begin{array}{r c l}
	\arraycolsep=1.4pt\def\arraystretch{2.5}
	c_1 &=&A_2+A_3+\cfrac{A_4\gamma}{x_0} \approx 397,2\\
	c_2 &=& A_2+A_3 \approx 117.5\\
	A_4 &=& 7.5\\
	A_1 &\approx& -17.7\\
	q(\sigma) &=& \cfrac{2c_2^3-9c_1c_2A_4}{27c_1^3}+\cfrac{A_1}{c_1}\sigma \approx 5,5.10^{-5}-0.04\sigma\\
	\Delta(\sigma) &=& q(\sigma)^2+\cfrac{4}{27}\left(\cfrac{3c_1A_4-c_2^2}{3c_1^2}\right)^3 \approx q(\sigma)^2-1,6.10^{-7},
\end{array}
\eeq
for the parameter values $\alpha = 1, \gamma=0.6$, $h_0=0.001$, $x_0=0.12$. Expression \eqref{xM_06} is valid as long as $\Delta(\sigma)\geq 0$, that is $\sigma>0.0114$.
\subsubsection{Expression of distance \boldmath $A_\sigma$ \unboldmath and \boldmath $A$ \unboldmath in the range \boldmath $0.885\leq\cfrac{\gamma}{\alpha}<1$}
\begin{figure}[http!]
\centering
\includegraphics[scale=0.74]{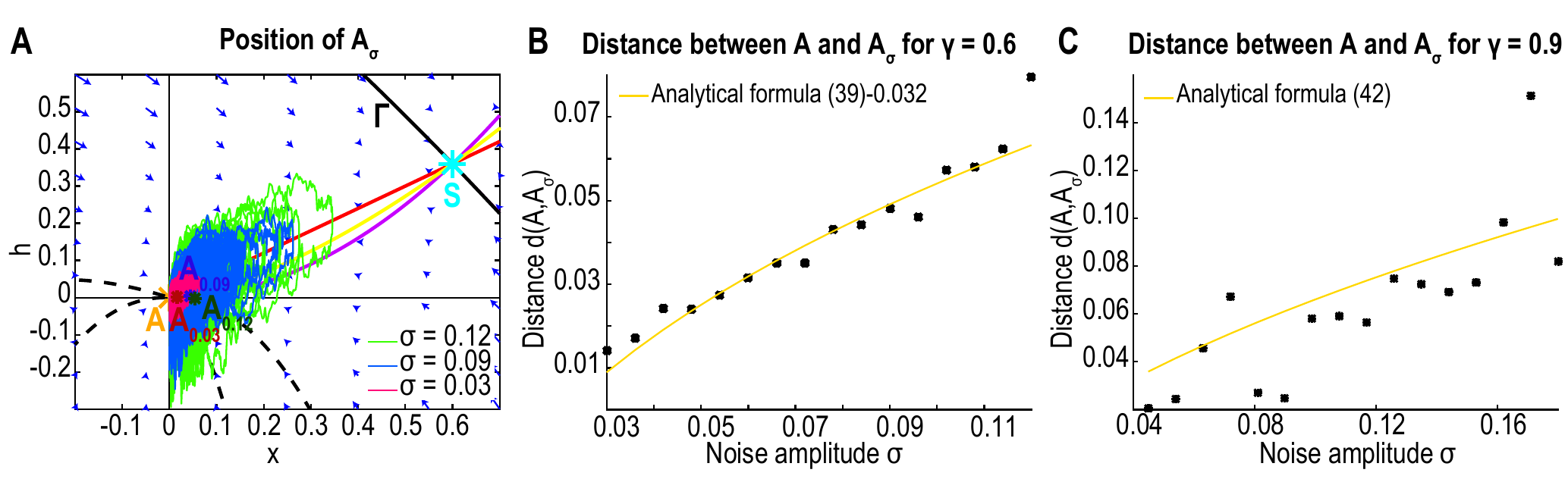}
\caption{\textbf{Influence of the noise amplitude on the peak distribution \boldmath $A_\sigma$ \unboldmath. }\textbf{A.} Stochastic trajectories simulated for $T=500s$ with three noise levels $\sigma = 0.03$ (pink), 0.09 (blue) and 0.12 (green) and the shifted peaks $A_\sigma$. \textbf{B.} Distance $d(A,A_\sigma)$ vs $\sigma$. Numerical simulations are obtained with a stopping time $T=800s$ per noise value (black stars), compared to the analytical formula \ref{xM_06} (yellow) minus a corrective offset $\hat{c}=0.032$. \textbf{C.} Distance $d(A,A_\sigma)$ vs $\sigma$. Numerical simulations are generated with a stopping time $T=800s$ per noise value, (black dots) compared to the analytical formula \ref{xM_09} (yellow).}\label{noise_Asigma}
\end{figure}

In the range $0.9\leq \frac{\gamma}{\alpha}<1$ we approximate \eqref{dpdxEq0} by the second order polynomial equation
\beq\label{dpdx_ordre2}
\cfrac{\gamma}{x_0}x^2+x+\cfrac{A_1 \sigma}{A_2+A_3+A_4}=0,
\eeq
the solution is
\beq\label{xM_09}
x_M(\sigma)=\cfrac{-1+\sqrt{1-\cfrac{4\gamma}{x_0}\cfrac{A_1 \sigma}{A_2+A_3+A_4}}}{2\gamma}x_0 \approx -0.56 + \cfrac{\sqrt{1+38.45\sigma}}{1.8},
\eeq
for the parameter values $\alpha=1$, $\gamma=0.9$ and $\sigma\geq 0$ (fig. \ref{phaseSpace_PDF}E cyan crosses and yellow curve).\\
To test the range of validity of our approximations for the position of $A_\sigma$, we ran simulations for $\sigma \in [0.03, 0.12]$ (fig. \ref{noise_Asigma}A $\sigma =0.12$, green, 0.09, blue and 0.03, pink). We compared the distance $d(A,A_\sigma)$ obtained from numerical simulations (\ref{noise_Asigma}B-C black stars) and the analytical formula \eqref{xM_06} (resp. \ref{xM_09}) (fig. \ref{noise_Asigma}B  (resp. C) yellow curve) for $\alpha=1$ and $\gamma=0.6$ (resp. $\gamma=0.9$). In the case $\gamma=0.6$, we added an offset in formula \eqref{xM_06} to minimize the absolute value of the difference between the analytical formula and the simulations:
\beq
\hat{c}=\min\limits_{c \in [0,0.1]}\sum\limits_{\sigma\in[0.03,0.12]}|x_M(\sigma)-c-x_{M,sim}(\sigma)|,
\eeq
where $x_M(\sigma)$ is defined by \eqref{xM_06} and $x_{M,sim}(\sigma)$ is the value obtained from the numerical simulations. Using a finite number of values, we directly found that $\hat{c}= 0.032$. This offset is probably due to the approximations we made for the derivation of formula \eqref{xM_06}. \\
To conclude we have found here analytical expressions (\ref{xM_06} and \ref{xM_09}) for the position of $A_\sigma$ and showed that these expressions approximate well the numerical simulations.
\section{Multiple re-entries and distributions of escape times and points}
In this section, we report a novel mechanism of stochastic escape from an attractor, based on multiple re-entry.
\subsection{The different steps of escape}\label{numerics_escape}
\begin{figure}[http!]
\centering
\includegraphics[scale=0.74]{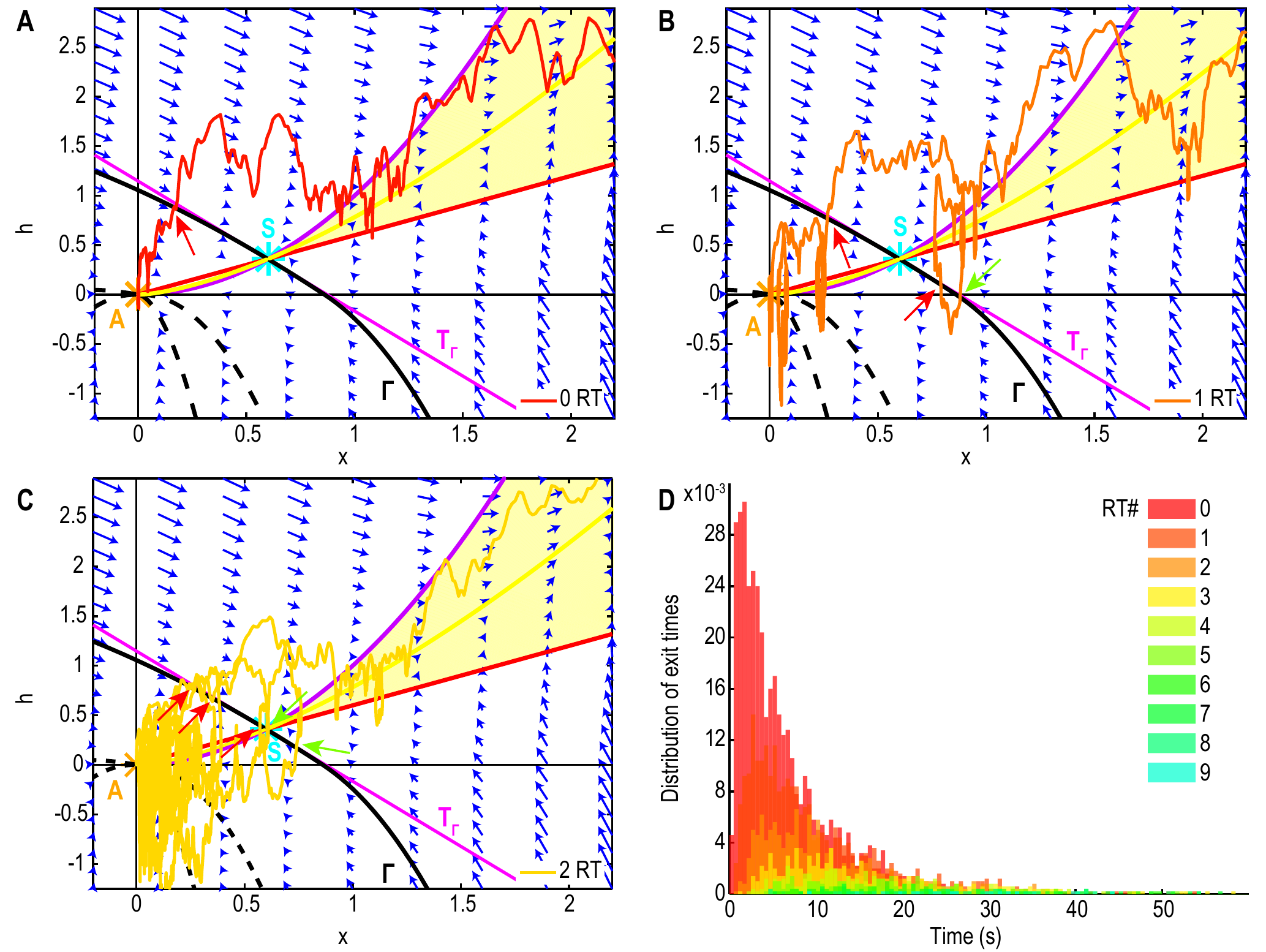}
\caption{\textbf{Recurrent exit pattern and contribution to the escape time. }\textbf{A-C.} Trajectory doing zero (resp. one, two) RT before escape (red, resp. orange, yellow) the red (resp. green) arrows indicate exit (resp. reentry) points. \textbf{D.}  Distribution of escape times from 5000 trajectories lasting T=300s ($\gamma =0.6$, $\alpha=1$ and $\sigma=0.78$) with the contribution of trajectories doing for each number of RT around $\Gamma$ before escaping (color gradient).}\label{RTtimes}
\end{figure}

The escape from the basin of attraction of point $A$ can be divided into three steps.
\begin{enumerate}
  \item Step 1: starting from the attractor $A$, trajectories fall into the basin of attraction of the shifted equilibrium $A_\sigma$. The duration of this step is almost immediate and can be neglected compared with the durations of the next steps 2 and 3.
  \item Step 2: trajectories fluctuate around the shifted equilibrium $A_\sigma$ until they reach the separatrix $\Gamma$ for the first time.
  \item  Step 3: trajectories cross $\Gamma$, exiting and reentering the basin of attraction several times before eventually escaping far away (fig. \ref{RTtimes}A-C).
\end{enumerate}
We will quantify these excursions occurring in step 3 by counting the number of round-trips (RT) across $\Gamma$. We first study these three steps numerically by simulating 5000 trajectories starting from $A$ and lasting $T=300s$ for $\sigma=0.78$. To obtain the distribution of exit times and points on the separatrix $\Gamma$ at each RT, we decided to replace $\Gamma$ by its tangent $T_\Gamma$ at $S$ (fig. \ref{RTtimes}A-C pink line). Indeed, the distribution of exit points peaks at a distance $O(\sqrt{\sigma})$ from $S$ \cite{BobrovskySchuss1982} and thus the difference between the separatrix and its tangent is of order 2. This approximation will allow us to use the analytical expression of the tangent $T_\Gamma$.\\
We then decompose the escape time in the first time to reach the separatrix $\Gamma$ plus the time spent doing successive excursions outside and inside the basin of attraction (fig. \ref{RTtimes}D, color gradient indicates the contribution of the trajectories doing a specific number of RT to the total distribution of escape times). This decomposition can be used to estimate the proportion of trajectories that escape at each RT and to evaluate the escape probability after crossing the separatrix as we will see below.
\subsection{First exit time and exit points distributions on the separatrix $\Gamma$}
\begin{figure}[http!]
\centering
\includegraphics[scale=0.74]{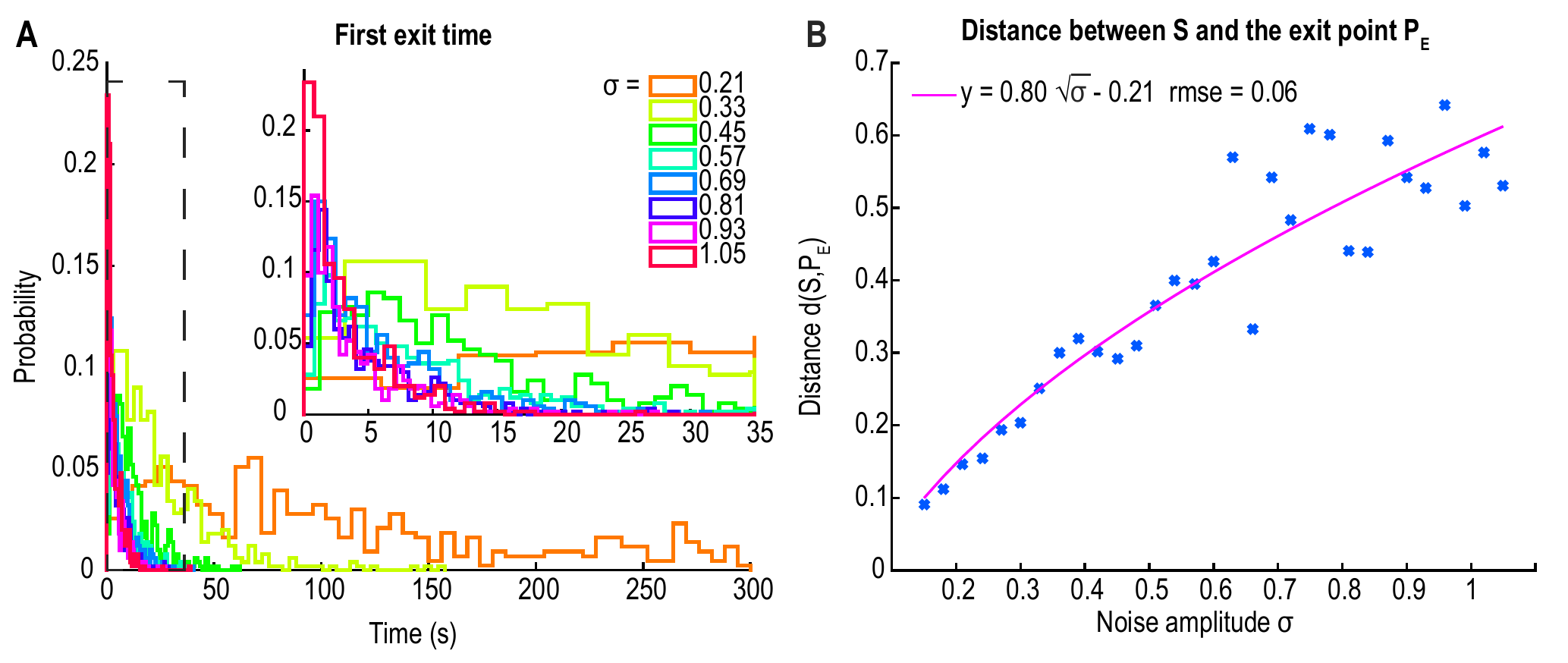}
\caption{\textbf{Influence of the noise amplitude on the first exit times and points. }\textbf{A.} Distribution of the first exit times with respect to the noise amplitude $\sigma$ with an inset on short times. \textbf{B.} Distance between the peak $P_E$ of the distribution of first exit points on the separatrix and the saddle-point $S$ with respect to the noise amplitude $\sigma$ for $\gamma=0.6$ and $\alpha=1$ with a square-root fit (magenta).}\label{variationSigma}
\end{figure}

We study here the influence of the noise on the distributions of first exit times and exit points: we simulate $N=2500$ trajectories starting at $A$ and lasting $T=300s$ for $\sigma \in [0.21,1.05]$. The first exit time can be very long for small values of $\sigma$ (fig. \ref{variationSigma}A, orange distribution for $\sigma = 0.21$, and light green for $\sigma = 0.33$) but becomes shorter with peaked distributions when $\sigma$ increases (dark green to red). The distribution of the first exit points is peaked and located on the left of the saddle-point $S$ (fig. \ref{S2}A purple for $\sigma = 0.78$). We found here that the distance $d(P_E,S)$ between the peak $P_E$ of this distribution and the saddle-point $S$ for $\sigma \in [0.15, 1.05]$ is of order $O(\sqrt{\sigma})$ (fig. \ref{variationSigma}B) in agreement with the classical theory \cite{BobrovskySchuss1982}. We further observed from numerical simulations that the density of exit points of the first trajectories that reenter the basin of attraction at least once follows a similar distribution as the one of first exit points for all trajectories (fig. \ref{S2}A green) indicating that there are no correlations between the position of the escape points and the phenomenon of reentry in the basin of attraction. Finally, the distribution of the first reentry points also peaks on the left of the saddle-point but spreads on both sides of $S$ (fig. \ref{S2}A red).
\subsection{Characterization of the mean escape time}\label{analytics_escape}
\begin{figure}[http!]
\centering
\includegraphics[scale=0.65]{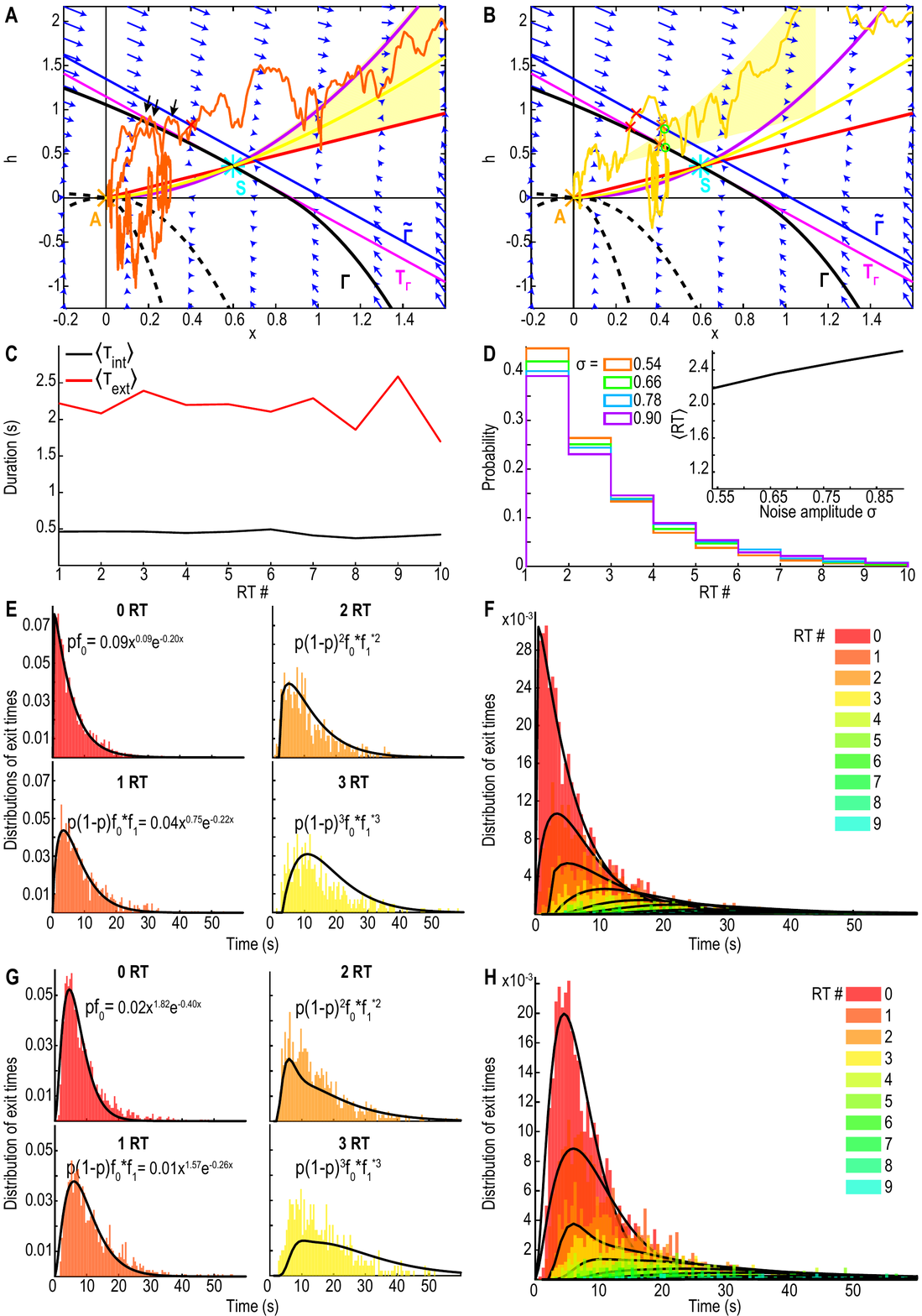}
\caption{\textbf{Characterization of escape times. }\textbf{A.} Trajectory escaping (red cross) after small brownian fluctuations at the separatrix (black arrows). \textbf{B.} Trajectory exiting (red crosses) after crossing $T_\Gamma$ (pink) and $\tilde{\Gamma}$ (blue), reentering (green circles) and reexiting (orange crosses) the basin of attraction before escaping. \textbf{C.} Mean time $\langle\tau_{ext}\rangle$ (resp. $\langle\tau_{int}\rangle$) spent outside (resp. inside) the basin of attraction at each RT. \textbf{D.} Distributions of the RT number before trajectories escape to infinity for $\sigma\in[0.54,0.90]$. Inset: mean RT number $\langle RT\rangle$ with respect to the noise amplitude. \textbf{E.} Distribution of exit times for trajectories doing 0 (upper left, resp. 1 (lower left), 2 (upper right), 3 (lower right)) RT before escape. \textbf{F.} Distribution of exit times with the contribution of each RT number (color gradient) with the analytical distribution \eqref{pdfExitTimes} (black) \textbf{G-H.} Application to the model \eqref{2Dsyst}}\label{escTimes}
\end{figure}

To obtain a general expression for the mean escape time, we use Bayes'law and condition by the RT numbers so that
\beq \label{realEscapeTime}
\langle\tau_{esc}\rangle = \sum_{k=0}^{\infty} \langle\tau | k \rangle P_{RT}(k),
\eeq
where $\langle \tau | k \rangle$ (resp. $P_{RT}(k)$) is the mean time (resp. probability) to return $k-$times inside the basin of attraction. Because the RT are independent events, the probability $\tilde{p}$ that a trajectory crossing the separatrix $\Gamma$ escapes does not depend on $k$, yielding
\beq
P_{RT}(k) = \tilde{p}(1-\tilde{p})^{k-1},
\eeq
thus
\beq\label{tEscSummed}
\langle\tau_{esc}\rangle = \langle\tau_{0}\rangle + (\langle\tau_{ext}\rangle+\langle\tau_{int}\rangle)\tilde{p}\sum_{k=1}^\infty k (1-\tilde{p})^{k-1} = \langle\tau_{0}\rangle + \frac{\langle\tau_{ext}\rangle+\langle\tau_{int}\rangle}{\tilde{p}},
\eeq
where $\langle\tau_{ext}\rangle$ (resp. $\langle\tau_{int}\rangle$) is the mean time spent outside (resp. inside) the basin of attraction during one RT. To avoid counting small Brownian fluctuations as RT inherent to the discretization (fig. \ref{escTimes}A black arrows), we added a second line $\tilde{\Gamma}$ at distance $\delta=0.25$ parallel to the separatrix (fig. \ref{escTimes}B blue line) and thus a trajectory is considered to have fully exited the basin of attraction once it has crossed both the tangent $T_\Gamma$ and $\tilde{\Gamma}$.\\
Using this procedure we estimated the probability $\tilde{p}(k)$ to escape after $k$ RT by counting the proportion of trajectories that reenter the basin of attraction at least once and we found using numerical simulations $\tilde{p}(1)\approx0.40$. We iterated this process for each RT until all trajectories had escaped to infinity and found that the probability $\tilde{p}(k)$ for $k \geq 1$ does not depend on $k$ (fig. \ref{S2}B), thus $\tau_{esc} \approx \tau_0 + 2.5(\tau_{ext}+\tau_{int})$. Finally, with the parameters $\alpha=1$, $\gamma=0.6$ and $\sigma=0.78$, numerical simulations show that $\langle\tau_0\rangle \approx 5s$ and $\langle\tau_{ext}\rangle+\langle\tau_{int}\rangle \approx 2.6s$ (fig. \ref{escTimes}C). \\
To conclude, the process of entering and exiting multiple times increases the mean escape time by a factor of 2.3. In addition we found, based on simulations for $\sigma\in[0.54,0.90]$, that the noise amplitude does not influence the number of RT before escape (fig. \ref{escTimes}D) and that trajectories perform 2.5 RT on average.
\subsection{Characterization of escape times distributions} \label{analytics_escape_distrib}
To determine the distribution of escape times, we condition the escape on the number of RT, so that
\beq\label{distrib}
P(\tau_{esc}<t)=\sum_{k=0}^{\infty} P(\tau^k<t|k) P_{RT}(k),
\eeq
where $P(\tau^k<t|k)$ is the probability distribution of escape times after $k$ RT. This probability is obtained by the $k$-th convolution of the distribution $f_1$ of times for trajectories exiting after a single RT with the distribution $f_0$ of escape times with 0 RT
\beq\label{pk}
P(\tau^k<t|k)=f_0(t)*f_1(t)^{*k},
\eeq
where $f(t)^{*k}=f(t)*f(t)*...*f(t)$, $k$ times. Thus \eqref{distrib} becomes
\beq\label{pdfExitTimes}
P(\tau_{esc}<t)=\sum_{k=0}^{\infty} f_0(t)*f_1(t)^{*k} \tilde{p}(1-\tilde{p})^{k}.
\eeq
To compare this formula with our numerical results, we decided to fit the distributions with
\beq\label{fitFi}
f_i(t)=c_i t^{\ds a_i} e^{\ds -\lambda_i t}, \text{ for } i=0,1, a_i\geq0 \text{ and } \lambda_i\geq0
\eeq
Using the Matlab fit function (fig. \ref{escTimes}C), we obtained for the distribution escape without any RT
\beq\label{F_0}
\tilde{p}f_0(t) = 0.09 t^{\ds 0.09}e^{\ds -0.20 t}
\eeq
and for the trajectories doing exactly one RT before escape
\beq\label{F1}
F_1(t)=\tilde{p}(1-\tilde{p})f_0*f_1(t) = 0.04 t^{\ds 0.75}e^{\ds -0.22 t}.
\eeq
To recover the distribution $f_1$ \eqref{F1}, we deconvolved numerically $F_1$ from $f_0$ \eqref{F_0} (fig. \ref{escTimes}E lower left). This procedure allows us to validate our approach by computing the distributions of 2 and 3 RT and comparing them with the empirical distributions \eqref{pk} (fig. \ref{escTimes}E upper and lower right). Finally, we decomposed the entire escape times distribution using \eqref{pdfExitTimes} to evaluate the contribution of each term (fig. \ref{escTimes}E-F).
\begin{center}
	\begin{tabular}{l l l}
		& Parameters & Values \\
		\hline
		$\tau$ & Time constant for $h$ & 0.05s\\
		$J$ & Synaptic connectivity & 4.21\\
		$K$ & Facilitation rate & 0.037Hz\\
		$X$ & Facilitation resting value & 0.08825\\
		$L$ & Depression rate & 0.028Hz\\
		$\tau_r$ & Depression time rate & 2.9s\\
		$\tau_f$ & Facilitation time rate & 0.9s\\
		$T$ & Depolarization parameter & 0\\
		\hline
	\end{tabular}
\label{tableParam}
\end{center}
\section{Further application and concluding remarks}
\subsection{Distribution of interburst durations} \label{application_2D}
The phase-space of system \eqref{2Dsyst}, restricted to the region $\{x \leq 0.5 \mbox{ and } h \leq 30\}$ is topologically equivalent to the phase-space of system \eqref{Drift}. It contains one attractor and one saddle-point and the stable manifold of the saddle-point defines the boundary of the basin of attraction (fig. \ref{application}A). Following the result of system \eqref{Drift}, trajectories fall into a basin of attraction centered around a shifted attractor towards the saddle-point $S$, as shown in fig. \ref{application}B, red (resp. blue, green) star for $\sigma=1$ (resp. $1.5$, $2.5$)). The shifted attractor position depends on the noise amplitude $\sigma$ (fig. \ref{application}C) and the escaping trajectories can return several times inside the basin of attraction before escaping far away (fig. \ref{application}D, one RT). We used formula \eqref{pdfExitTimes} to fit the distribution of exit times (fig. \ref{application}E-F) and obtained, for $\sigma = 6$ the distribution of exit with no return
\beq
f_0(t) = 0.02 t^{\ds 1.82}e^{\ds -0.40 t}
\eeq
and the distribution of loop is
\beq
f_1(t) = 0.01 t^{\ds 1.57}e^{\ds -0.26 t}.
\eeq
Finally, using numerical simulations, we estimated the escape probability $\tilde{p} \approx 0.37$ by generating trajectories starting from the attractor A and counting the fraction that fully escaped far away vs those that did return inside the basin of attraction (fig. \ref{S3}A).  The mean escape time is given by formula \eqref{tEscSummed}: \beq
\langle\tau_{esc}\rangle \approx \langle\tau_0\rangle + 2.7(\langle\tau_{ext}\rangle+\langle\tau_{int}\rangle).
\eeq
Using parameters of table \ref{tableParam}, we obtain $\langle\tau_0\rangle \approx 4.35s$ and $\langle\tau_{ext}\rangle+\langle\tau_{int}\rangle\approx 2.6s$ (fig. \ref{S3}B). We conclude that returns to the attractors increases the escape time from 4.35 to 11.37 leading to an increase by a factor 2.6. Moreover, the number of RT before escape does not depend on the noise amplitude (section \ref{analytics_escape} and fig. \ref{S3}C) and trajectories generate in average 2.7 RT before escape (fig. \ref{S3}C, inset). Finally, this escape mechanism could explain long interburst durations occurring in excitatory neuronal networks without the need of adding any other refractory mechanisms.  Indeed, the present computations can be used to study the neuronal interburst dynamics modeled in the mean-field approximation by depression-facilitation equations \cite{Tsodyks1997,daoduc2015}. We also provided a possible explanation for the long interburst durations observed in neuronal networks reported in \cite{Rouach_CxKO}.\\
\begin{figure}[http!]
\centering
\includegraphics[scale=0.74]{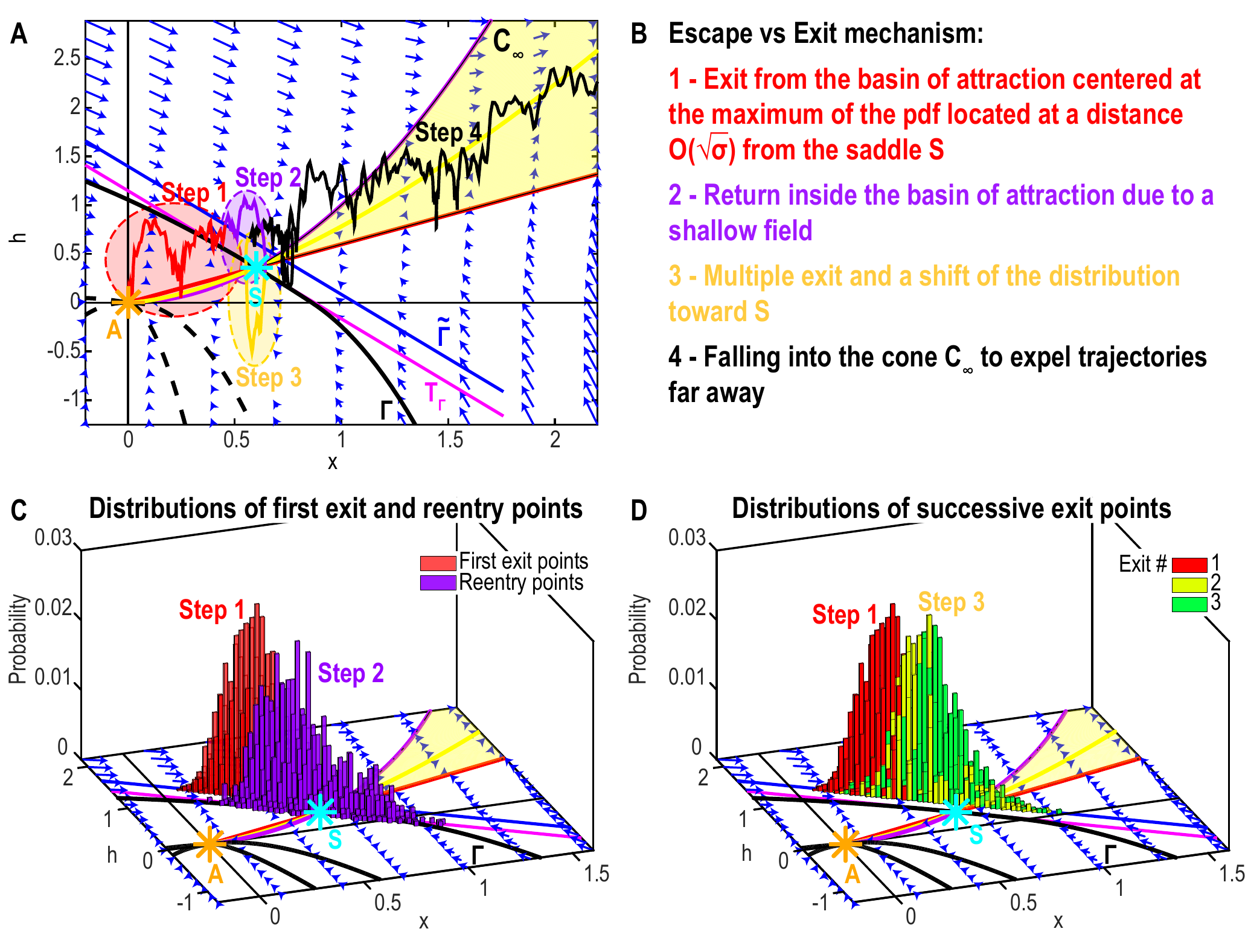}
\caption{\textbf{Recurrent escape mechanism. }\textbf{A.} Trajectories exiting the basin of attraction (red), re entering (purple) and exiting again (yellow) before eventually escaping to infinity (black) after reaching the cone $C_\infty$ (yellow surface). \textbf{B.} Schematic of escape process divided in four steps. \textbf{C.} Distribution of the first exit (resp. reentry) points (red, resp. purple) on the separatrix $\tilde{\Gamma}$. \textbf{D.} Distributions of successive exit points (first exit, red, second yellow, third green) on $ \tilde{\Gamma} $.}\label{recap}
\end{figure}
To conclude this article, motivated by finding a possible mechanism that generates long interburst intervals, we examined a family of stochastic dynamical systems perturbed by a small Gaussian noise. The dynamics exhibits specific properties such as peaks of the pdf inside the basin of attraction shifted compared to the attractor. In addition, escaping the basin is characterized by multiple reentries inside the attractor. We computed the position of this shifted attractor using WKB approximation and we derived algebraic formulas to link the position to the noise amplitude $\sigma$ (formulas \ref{xM_06} and \ref{xM_09}). We also computed the escape time, decomposed into the time to reach the boundary of the basin of attraction plus the time spent going back and forth through the separatrix (formulas \ref{tEscSummed} and \ref{pdfExitTimes}). Finally, we emphasize the generic conditions associated with this escape dynamics into four conditions:
\begin{enumerate}
\item The distribution of exit points peaks at a distance $O(\sqrt{\sigma})$ from the saddle-point (generically satisfied \cite{BobrovskySchuss1982}).
\item The shallow field near the separatrix allows the trajectories to reenter the basin of attraction with high probability.
\item The peaks of the successive exit points distributions drift towards the saddle-point $S$ (fig. \ref{recap}C).
\item When trajectories enter the cone $C_{\infty}$ (yellow surface in fig. \ref{recap}A-B) where the field increases, they eventually escape to infinity.
\end{enumerate}
{\bf Acknowledgements}
L. Zonca has received support from the FRM (FDT202012010690). This project has received funding from the European Research Council (ERC) under the European Union’s Horizon 2020 research and innovation program (grant agreement No 882673).
\subsection*{Appendix Figures}
\renewcommand\thefigure{A.\arabic{figure}}
\setcounter{figure}{0}
\begin{figure}[http!] 
\centering
\includegraphics[scale=0.74]{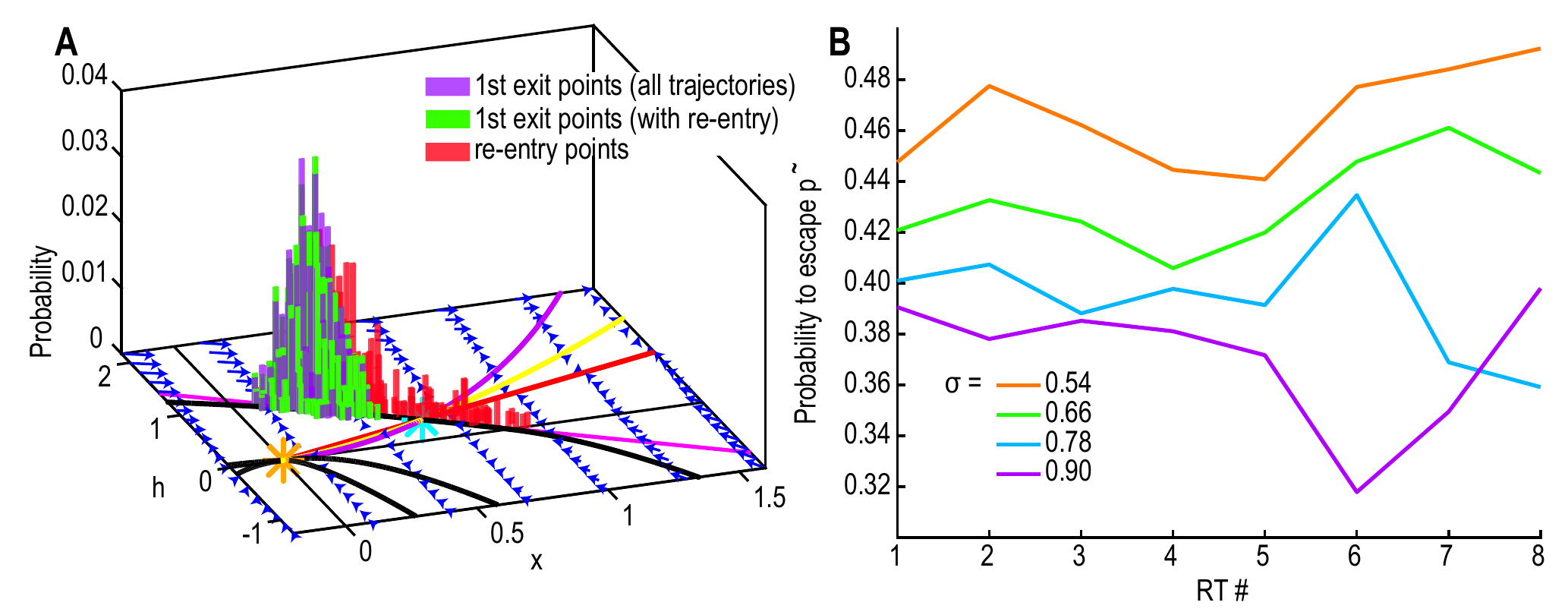}
\caption{{\bf  Distributions of the first exit points located on the separatrix $\Gamma$ }\textbf{A.} for all trajectories (purple) and first exit for re-entering trajectories (green, 60\% of trajectories), reentry points (red)  for $\sigma=0.78$. \textbf{B.} Probability to escape after exiting the basin of attraction for the $k$-th time, $\tilde{p}(k)$ estimated by the proportion of trajectories that reenter the basin of attraction after $k$ RT for noise amplitudes $\sigma \in [0.54,0.90]$.} \label{S2}
\end{figure}
\begin{figure}[http!] 
\centering
\includegraphics[scale=0.74]{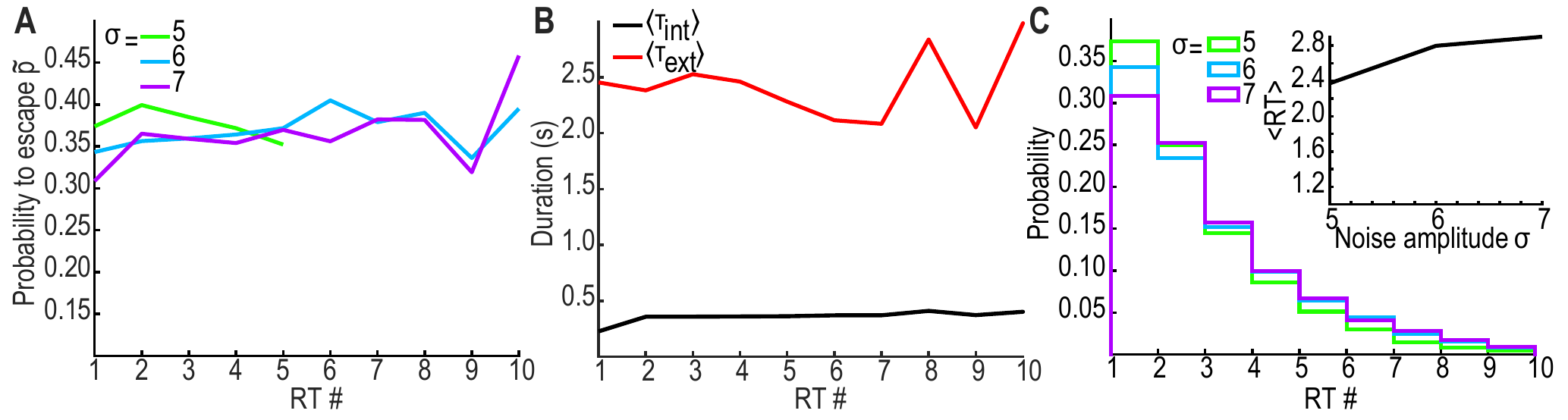}
\caption{{\bf Probability $\tilde{p}$ to escape after exiting the basin of attraction} \textbf{A.}  Probability $\tilde{p}$  vs the RT number for $\sigma \in [4,7]$, with a linear fit. \textbf{B.} Mean times  $\langle\tau_{ext}\rangle$ (red) and $\langle\tau_{int}\rangle$ (black) vs the RT number. \textbf{C.}  Distributions of RT numbers before escape for $\sigma\in[4,7]$. Inset: mean RT number with respect to $\sigma$} \label{S3}
\end{figure}
\newpage
\normalem
\bibliographystyle{ieeetr}
\bibliography{biblio2}
\end{document}